\documentclass[3p,twocolumn]{elsarticle}
\usepackage{amssymb}
\usepackage{epsfig}
\usepackage{pifont}
\usepackage{booktabs}

\journal{Astroparticle Physics}

\usepackage{todonotes}
\let\oldtodo\todo
\renewcommand{\todo}[1]{%	
	\oldtodo[inline]{#1}%
}

\newsavebox\CBox  
\newcommand\hcancel[2][0.5pt]{%
  \ifmmode\sbox\CBox{$#2$}\else\sbox\CBox{#2}\fi%
  \makebox[0pt][l]{\usebox\CBox}%  
  \rule[0.5\ht\CBox-#1/2]{\wd\CBox}{#1}}

\usepackage{soul}

\begin{document}

\begin{frontmatter}
\title{Adaptive Kernel Density Estimation for Improved Sky Map Computation in Gamma-Ray Astronomy} 

\author[uibk]{M.~Holler}
\ead{markus.holler@uibk.ac.at}

\author[uibk]{T.~Mitterdorfer}
\author[uibk]{S.~Panny}

\address[uibk]{Universit\"at Innsbruck, Institut f\"ur Astro- und Teilchenphysik}
\begin{abstract}
We introduce an alternative method for the calculation of sky maps from data taken with gamma-ray telescopes. In contrast to the established method of smoothing the 2D histogram of reconstructed event directions with a static kernel, we apply a Kernel Density Estimation (KDE) where the kernel size of each gamma-ray candidate is related to its estimated direction uncertainty. Exploiting this additional information implies a gain in resulting image quality, which is validated using both simulations and data. For the tested simulation and analysis configuration, the achieved improvement can only be matched with the classical approach by removing events with lower reconstruction quality, reducing the data set by a considerable amount.
\end{abstract}


\end{frontmatter}

\section{Introduction}
\label{sec_intro}

Imaging Atmospheric Cherenkov Telescopes (IACTs) measure the Cherenkov light from electromagnetic particle showers to reconstruct the direction and energy of very-high-energy (VHE; $E \gtrsim 100\,$GeV) gamma-rays. Despite many advancements in the last decades, the resulting directional reconstruction quality remains poor compared to most other wavelength regimes, mostly owing to the statistical scatter of the shower particles. The resulting point spread function (PSF) exhibits $68\%$ containment radii at the level of $R_{68} \approx 0.05-0.1^{\circ}$, depending on the observation and detector conditions as well as the analysis settings. Nevertheless, numerous extended sources with partly complex morphologies have already been observed, especially in the Galactic Plane \cite{2018_HGPS}. However, the typically small number of detected gamma-rays in combination with a generally small signal-to-background ratio necessitates the usage of smoothing algorithms for visualisation purposes when creating maps. This convolution of intrinsic detector response (PSF) and additional smearing results in a broadening of the actual morphological features in the final sky maps. 

The current standard for generating sky maps in IACT gamma-ray astronomy is to fill the reconstructed event directions into a two-dimensional histogram with a bin width $b \ll R_{68}$ for both axes. This raw histogram is subsequently smoothed, often with a top-hat kernel (also called oversampling), where the kernel width $w$ is usually chosen to be of the order of $R_{68}$. The whole approach is straightforward and computationally efficient, but the fixed kernel width ignores that the PSF is not a constant but actually varies from event to event. IACT observations of a given sky region almost always correspond to a diverse data set in terms of observational conditions like zenith angle and atmospheric transparency as well as telescope and camera conditions \citep{2006_HofmannResolution,2006_HessCrab,2012_MAGICPerformance}. 
Even during a given observation and for gamma-rays from the same source, parameters like the distance of the shower axis to the telescope array or the energy of the primary gamma-ray strongly influence the directional reconstruction quality. This effect is notably expected to increase even more for upcoming next-generation instruments like the Cherenkov Telescope Array (CTA, \citep{2019_ScienceWithCTA}), where a greater amount of telescopes as well as different telescope types will result in a data sample with a larger range of event direction uncertainties.
All in all, characterising the PSF of an observed sky region with only one number when creating a sky map is a strong simplification. The situation is ameliorated when separating the data sample into events of different reconstruction quality, as is common standard in the case of data from \textit{Fermi}-LAT  \citep{2013_FermiPass8}. Modern analysis frameworks in gamma-ray astronomy, such as \textit{gammapy} and \textit{Fermipy} \citep{2017_gammapy,2017_Fermipy}, allow combined likelihood analyses of the independent data sets corresponding to the different event types. The separate treatment of events of different quality accompanied by their corresponding Instrument Response Functions leads to an improved sensitivity, which is why this approach has also already been suggested for CTA \citep{2021_Hassan}.

Although the actual directional uncertainty per event is naturally only known for simulated gamma-rays, likelihood-based reconstruction techniques allow calculating estimates of the uncertainty of each fit parameter, including the direction, from the fit correlation matrix \cite{2009_Model}. This direction uncertainty corresponds to an information gain which can be used for better imaging.

Here we introduce a novel method for the generation of sky maps in gamma-ray astronomy via Kernel Density Estimation (KDE). By using the event-wise, estimated direction uncertainty as input for the kernel width, all available directional information from the reconstruction process is retained for the sky-map computation. The morphology of gamma-ray sources in the resulting maps are correspondingly better resolved, to a level that is only achievable with conventional methods when removing events with lower reconstruction quality. In contrast to that, our approach provides a high resolution whilst keeping the full data sample. Although we only use IACT data for demonstration, the method is expected to be equally applicable to data from other types of gamma-ray telescopes whenever the direction uncertainty for a given candidate event can be estimated.

In Section~\ref{sec_methods}, we discuss the relevant concepts, namely the estimated direction error $\delta$ as well as the applied KDE. These are tested and compared to the classical histogram smoothing on both simulations and data in Section~\ref{sec_results}, followed by concluding remarks in Section~\ref{sec_conclusions}.

\section{Methods}
\label{sec_methods}

In this section, the methodology of the applied approach is presented.

\subsection{Direction Error}
\label{subsec_direrr}

The estimated direction uncertainty $\delta_i$, henceforth called direction error, for a given reconstructed event $i$ was introduced by \cite{2009_Model} for IACTs. The underlying reconstruction mechanism described there is based on a log-likelihood fit of templates of the shower intensity to the actual data. By varying the input parameters of the shower templates, the log-likelihood value is minimised to obtain the best-fitting properties of the primary particle, including the two spatial coordinates. The correlation matrix describes the behaviour of the log-likelihood profile around the obtained minimum, allowing to calculate estimated uncertainties of each fit parameter. $\delta_i$ then corresponds to the combination of both estimated directional uncertainties, a reduction which is justified when the PSF can be considered radially-symmetric, as it is mostly the case for the H.E.S.S. array of IACTs \cite{2020_CenA_Hess,2020_RWS}. We will only discuss this case in the following, but note that our method is expected to also be applicable to asymmetric PSFs when utilizing both directional error estimates separately.

\begin{figure*}
\includegraphics[width=\textwidth]{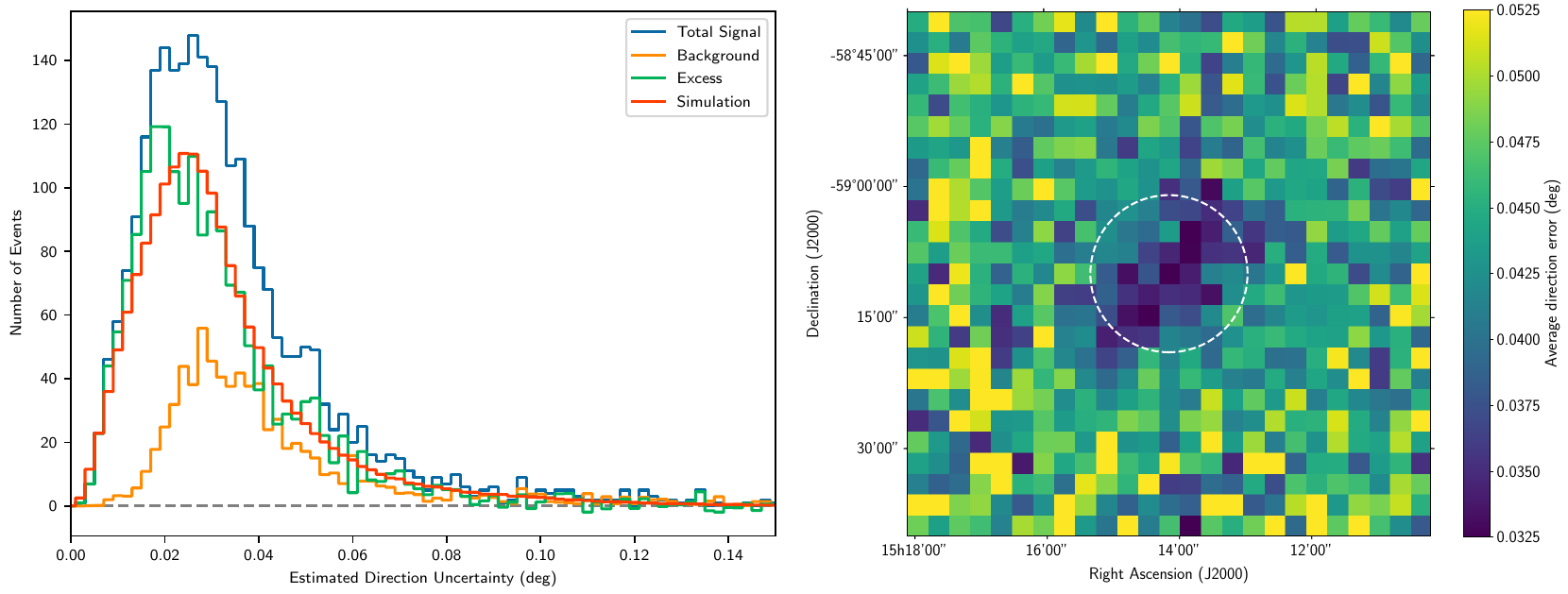}%
\caption{Illustrations of the direction error $\delta$ around the extended VHE gamma-ray source MSH15-5\textit{2}, obtained from 32 hours of observations with the H.E.S.S. array. The data were reconstructed with the \textit{Model} analysis \citep{2009_Model}, and only events that passed the \textit{standard} analysis cuts are shown. The resulting event sample exhibits an energy threshold of around $0.3\,$TeV and reaches to energies beyond $10\,$TeV. \textit{Left panel:} Distribution of $\delta$ within a circular region with radius $0.15^{\circ}$ around the source. The background histogram was obtained with the \textit{reflected-region} method \cite{2007_Berge}. The red distribution corresponds to simulated events specifically generated for this data set, normalised to the number of excess events. % last sentence fine?
\textit{Right panel:} Sky map showing the average value of $\delta$ for each position. The dashed white circle denotes the extraction region for the \textit{left panel}.}
\label{Fig_DirErr}
\end{figure*}
An example of the behaviour of the direction error near an actual, extended VHE gamma-ray source is shown in Fig.~\ref{Fig_DirErr}, containing two illustrations of $\delta$ around the Pulsar Wind Nebula (PWN) MSH15-5\textit{2} as measured with the four smaller IACTs of H.E.S.S. \cite{2005_HESSMSH}. The \textit{left panel} shows the distribution of $\delta$ within a region that contains most of the gamma-ray emission of the source, corresponding to events with different telescope multiplicity, energy, and other parameters that influence the reconstruction. The broad range of $\delta_i$ values indicates the large event-to-event variability of the directional reconstruction accuracy that was already mentioned in Sec.~\ref{sec_intro} and which builds the foundation of improvement of this work. The figure also includes an estimation of the residual background in that region that passes the analysis cuts, estimated using the \textit{reflected-region} method \cite{2007_Berge}. These background events mostly correspond to Cherenkov emission from hadronic showers initiated by charged cosmic rays as well as that of electromagnetic showers from cosmic electrons. As can be seen, the direction error values of the background events are generally larger than that of the excess events from the source, implying that the direction reconstruction is more precise for the latter. To check whether the observed behaviour can be reproduced, simulations of gamma-rays were generated for this data set using the approach described in \cite{2020_RWS}. Apart from small discrepancies at lower values of $\delta$, the corresponding distribution of the direction error agrees well with the one of the excess emission. The simulations were furthermore used to check the correlation of $\delta$ with the actual offset of reconstructed and true direction. On average, $\delta$ underestimates the actual offset by a factor of around $1.5$. Both variables exhibit a Pearson correlation coefficient of $0.455$, proving that the estimated direction error is indeed suited to identify events with good actual direction reconstruction accuracy (as evaluated for these MC simulations by calculating the offset between true and estimated directions). This correlation has already been exploited for numerous H.E.S.S. publications by only allowing events with a certain maximum value of $\delta$ (see, e.g., \cite{2020_CenA_Hess,HESS_RXJ1713,2019_CrabHess}).

The overall larger values of $\delta$ for background events are also clearly visible in the \textit{right panel} of Fig.~\ref{Fig_DirErr}, which displays the average direction error for the Field of View (FoV) around MSH15-5\textit{2} in a sky histogram. Despite the background contamination, the source distinctly stands out with its lower average direction error, highlighting the usefulness of this quantity for improved mapping techniques.

\subsection{Adaptive Kernel Density Estimation}
\label{subsec_kde}

In general, the aim of visualising the discrete data from IACTs is to highlight the observed sources in a way that best approximates the unknown distribution of the underlying emission.
As it was shown in Sec.~\ref{subsec_direrr}, the event-wise direction error $\delta_i$ contains additional information about the reconstruction accuracy, allowing to achieve better results in this regard. To properly exploit this information gain without making assumptions about the source morphology, a non-parametric density estimator that can handle event-wise accuracy is needed.
These requirements are fulfilled by the KDE method which was first introduced by \cite{1956_Rosenblatt} and \cite{1962_Parzen}. Following this approach, an estimate of an unknown probability density function $f(x)$ in one dimension can be calculated from a set of independent measurements ${x_1,x_2,...,x_n}$ as follows:
\begin{equation}
\label{eq_kde_orig}
\hat{f}( x ) = \frac{1}{nh}\sum_{i=1}^{n} K\left( \frac{x-x_i}{h} \right) \, ,
\end{equation}
where $K(x)$ denotes a suitable kernel function whose width is influenced by the smoothing parameter $h$. In words, a KDE provides an estimate of the underlying probability density by assigning each event a smooth function that is centred on its measured position $x_i$. While the choice of $h$ is often non-trivial because of unknown knowledge of the underlying scattering uncertainty of $x_i$, it should in our use case be directly related to the extent of the PSF. Exploiting the previously mentioned correlation between $\delta_i$ and the true direction offset as well as generalising Eq.~\ref{eq_kde_orig} to allow a non-constant smoothing parameter, we get the refined formulation
\begin{equation}
\label{eq_kde_refined}
\hat{f}\left( x \right) = \frac{1}{n}\sum_{i=1}^{n} \frac{1}{a \delta_i} K\left( \frac{x-x_i}{a \delta_i} \right) \, ,
\end{equation}
where $a$ corresponds to a constant scaling parameter. Letting the smoothing parameter vary from event to event as well as using $\delta_i$ implies a morphological information gain for the final image without having to reduce the data sample. Our final formulation of an adaptive KDE that is a suitable alternative to conventional two-dimensional event histograms in IACT gamma-ray astronomy then looks as follows:
\begin{equation}
\label{eq_kde_final}
\hat{F}( \mathbf{r} ) = \sum_{i=1}^{n} \kappa_i \left( \frac{\vartheta\left( \mathbf{r},\mathbf{r}_i  \right)}{a \delta_i} \right) \, ,
\end{equation}
with the angular distance $\vartheta \left( \mathbf{r},\mathbf{r}_i  \right)$ between the evaluated direction $\mathbf{r}$ and the reconstructed event direction $\mathbf{r}_i$. $\kappa_i$ corresponds to the normalised kernel function per event, however the overall normalisation factor $1/n$ is now omitted in contrast to Eq.~\ref{eq_kde_refined} to preserve the integral of the overall resulting count map. 

As visible in the \textit{left panel} of Fig.~\ref{Fig_DirErr}, the $\delta$ distributions of both simulated and actual data exhibit values down to almost $0.00^{\circ}$. However, the actual resolution limit for a H.E.S.S.-like instrument is rather at the order of around $1\,$arcmin or slightly below, depending on the photon energy \cite{2006_HofmannResolution}. We therefore apply a minimum value of $\delta_{\mathrm{min}} = 0.017^{\circ}/a$, which is taken instead of $\delta_i$ for those events where $\delta_i < \delta_{\mathrm{min}}$.

While there is no common standard in IACT gamma-ray astronomy for the size of the smoothing radius, the typical choice is a value around the size of the intrinsic PSF. Following this logic, we set $a = 1.5$ to correct for the discrepancy between estimated and actual direction error as observed on our example data set (see Sec.~\ref{subsec_direrr}). 

The outlined approach is realised by defining a finely binned two-dimensional grid\footnote{The code is available upon request from the authors.}, similar to the two-dimensional histograms that are traditionally used. For a given event $i$, the angular offset of the center of each grid point to $\mathbf{r}_i$ is calculated, allowing, together with the associated $\delta_i$, to evaluate the kernel function. The normalisation of the event-wise kernel is conducted numerically by dividing the obtained values by their sum. By avoiding analytical integration, this approach ensures full flexibility regarding the choice of the kernel function. We are therefore going to demonstrate our approach by applying both Gaussian as well as top-hat kernel functions in the upcoming section.

\section{Results} 
\label{sec_results}

In the following, we are going to demonstrate the previously introduced method of adaptive KDE and compare it to the classical approach, corresponding to a two-dimensional event histogram which is smoothed with the same kernel function. To minimise unwanted binning effects, we choose a bin size of only $0.001^{\circ}$ along both dimensions and use the same spacing for the evaluation grid of the KDE. For such a fine binning, the smoothing of a two-dimensional event histogram notably becomes almost identical to a static, i.e. non-adaptive, KDE with same kernel function and size\footnote{This was extensively checked and becomes mathematically obvious for infinitesimally small bin sizes.}. To ensure a fair comparison between both approaches, we consistently use the mean of the distribution of $\delta$ values after applying the restriction on the minimum value, multiplied by $a = 1.5$, as the kernel width for the histogram smoothing. 
\begin{figure*}
\includegraphics[width=\textwidth]{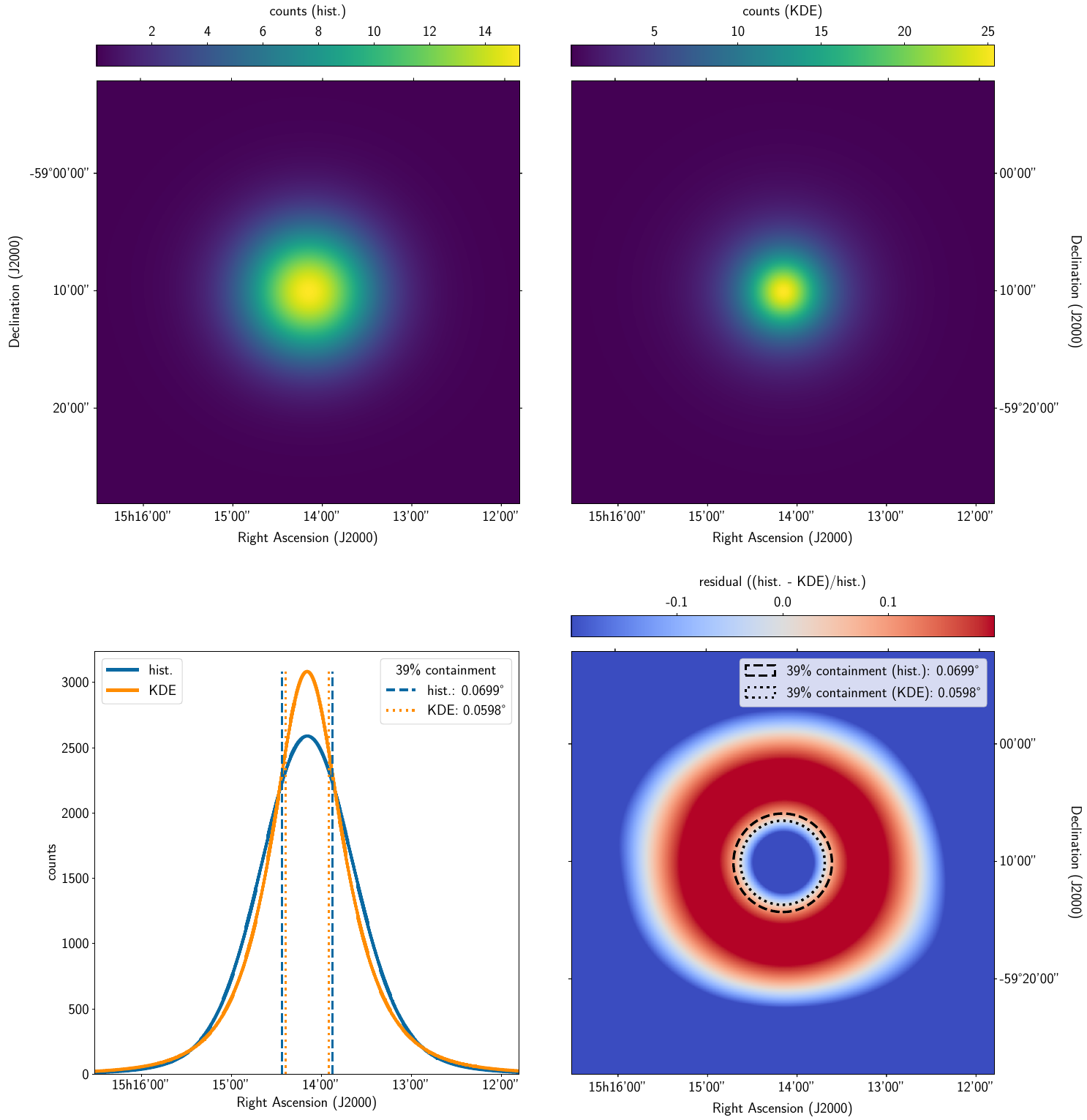}%
\caption{Comparison of methods to build sky maps applied to a simulated point source at the centre of the extraction region used in Fig.~\ref{Fig_DirErr}. \textit{Upper left:} Two-dimensional counts histogram, smoothed with a Gaussian kernel of $0.0504^{\circ}$ width, corresponding to the mean of the distribution of values of $\delta_i$ of the underlying events, scaled with $a = 1.5$. \textit{Upper right:} Result of the adaptive KDE for the same event sample, using a Gaussian kernel with event-wise width $a\cdot \delta_i$. \textit{Lower left:} Projections of the respective map content onto the right ascension axis. \textit{Lower right:} Relative residual map of the two approaches, including circles that illustrate the apparent $39\%$ containment inclusions for both.}
\label{Fig_SimPSF}
\end{figure*}

As a first application, both methods are compared using the simulated point-source events that were already used in Section~\ref{subsec_direrr}. The corresponding event maps are shown in Fig.~\ref{Fig_SimPSF}, clearly illustrating the improvement of the adaptive KDE (\textit{upper right panel}) over the standard histogram-smoothing approach (\textit{upper left panel}) by exhibiting a sharper peak due to the well-reconstructed events and more pronounced tails corresponding to events with less certain direction reconstruction. The latter is readily visible in the corresponding residual map (\textit{lower right panel}) as well as the projections onto the right ascension axis (\textit{lower left panel}). The smaller containment radius of the adaptive KDE method directly implies an improved imaging quality for maps that are generated with our approach while keeping the full event sample. When using the classical histogram-smoothing approach, a corresponding improvement is achievable by keeping only events with good reconstruction accuracy. In the case of this data set, the $39\%$ containment radii in the maps for adaptive KDE and smoothing converge in case only events with $\delta < 0.0225^{\circ}$ are kept for the latter. This additional selection comes at the cost of reducing the event sample by $69\%$. It has to be noted that, owing to the different shape of the distributions, this number highly depends on the applied containment ratio. For example, the $68\%$ containment radii of histogram smoothing and adaptive KDE methods can be matched by applying a cut of $\delta < 0.0540^{\circ}$ for the classical approach, where only $12\%$ of the events are disregarded in this case.

\begin{figure*}
\includegraphics[width=\textwidth]{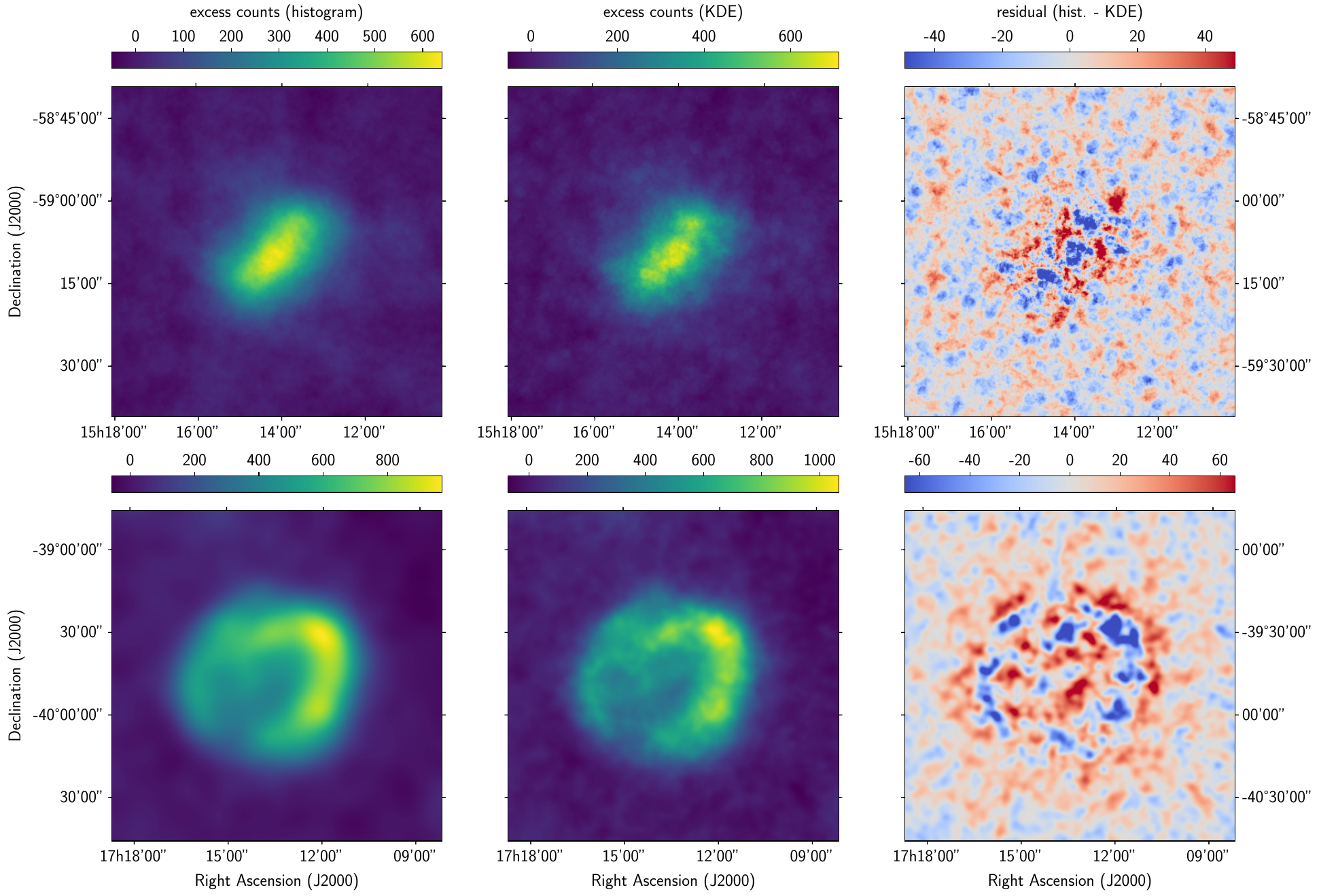}%
\caption{Illustration of histogram-smoothing (\textit{left panels}) and adaptive KDE approach (\textit{middle panels}), applied to H.E.S.S. data of MSH15-5\textit{2} (\textit{upper row}, using a top-hat kernel function) and RX J$1713.7$-$3946$ (\textit{lower row}, Gaussian kernel). All four maps are background-subtracted and renormalized with a correlation radius that corresponds to that of the histogram smoothing, enabling a direct pixel-by-pixel comparison of the methods. The \textit{right panels} contain the absolute residual maps of both approaches, i.e. the residual obtained when subtracting the excess map of the adaptive-KDE approach from the one obtained the classical way.}
\label{Fig_DataComp}
\end{figure*}
Following the validation on simulated data, both approaches are compared in the \textit{upper row} of Fig.~\ref{Fig_DataComp} on the data set of MSH15-5\textit{2} that was already used in Section~\ref{sec_methods}, applying a top-hat kernel function in this case. The final maps have been background-subtracted using the ring-background technique \citep{2007_Berge}. While both imaging methods are provided with the same data sample, exploiting the event-wise direction error with the adaptive KDE clearly leads to a sharper resulting image compared to the classical smoothing approach, as is also visible in the residual map (\textit{right panel}). The improvement is furthermore clearly visible in the \textit{lower row} of Fig.~\ref{Fig_DataComp}, showing a corresponding comparison using H.E.S.S. observations of the supernova remnant (SNR) RX J$1713.7$-$3946$ \citep{HESS_RXJ1713}, however applying a Gaussian kernel. The result from the adaptive KDE again yields a higher image quality than that obtained with the constant smoothing, revealing structures that are otherwise smeared out.  

By construction, the computation time of the adaptive KDE method depends both linearly on the number of events to be processed and that  of spatial bins (and therefore quadratically on the angular bin extent per axis for a fixed FoV size). To give an example, processing the $3.2\cdot 10^5$ events of our data sample from the direction of RX J$1713.7$-$3946$ using a FoV with $2^{\circ}$ edge length and $0.02^{\circ}$ bin size takes around $0.5\,$min on a standard computer equipped with an M1 ARM processor. Although this is considerably longer than the processing time needed to apply a standard 2D histogram smoothing, it is still feasible also for considerably larger data sets.

\section{Conclusions}
\label{sec_conclusions}

Here we have introduced a novel method for the generation of sky maps in IACT gamma-ray astronomy. It improves upon the traditional smoothing approach by utilizing the event-wise direction uncertainty estimate $\delta_i$ as provided by the \textit{Model} event reconstruction method \citep{2009_Model}. Instead of a static smoothing width, we employ an adaptive KDE, with an event-wise kernel width that is set proportional to $\delta_i$. This usage of additional information leads to better-resolved images, which has been demonstrated on both simulations and data. In addition to highlighting regions where the reconstruction accuracy is better than on average, the adaptive approach in turn also assigns a lower accuracy to events with poor accuracy, avoiding false features where the localization precision is worse than on average. In the case of a simulated point source, the adaptive KDE approach leads to a sharper result that can only be matched with the classical static smoothing by cutting on $\delta_i$ and reducing the event sample considerably. Naturally, the adaptive KDE will by construction perform even better when provided with this reduced data sample. As expected from these findings, the adaptive KDE method also leads to better results when applied to real data, which has been exemplarily demonstrated on data sets of both the PWN MSH15-5\textit{2} and the SNR RX J$1713.7$-$3946$. All our demonstrations were conducted with Gaussian and top-hat kernels, but the computational realisation allows to easily implement other suitable kernel functions. 

The grade of improvement of the adaptive KDE method strongly depends on the quality of the direction-error estimation. Studies using our simulation sample revealed an acceptable correlation coefficient of $0.455$ between estimated and actual direction error. While this value justifies the usage of $\delta$ in its current state for improved imaging methods, it also indicates potential room for improvement regarding the error estimation of the event reconstruction, which would in turn directly improve the result of the adaptive KDE. 
Furthermore, it has to be noted that our presented method is also expected to work with other variables that show a correlation with the actual direction uncertainty. It is therefore widely applicable in the field and not just restricted to data where $\delta_i$ is provided, posing a natural alternative to classical smoothing whenever better image quality is desired.

\section*{Acknowledgements}

We thank Stefan Wagner and Mathieu de Naurois, spokesperson and deputy spokesperson of the H.E.S.S. Collaboration, as well as Nukri Komin, chair of the H.E.S.S. Collaboration Board, for allowing us to employ the data used in this publication. We are furthermore grateful to the anonymous referees for their constructive feedback.

This research made use of \textit{gammapy}\footnote{https://www.gammapy.org}, a community-developed core Python package for TeV gamma-ray astronomy \citep{gammapy:2017}. This work made use of \textit{Astropy}:\footnote{http://www.astropy.org} a community-developed core Python package and an ecosystem of tools and resources for astronomy \citep{astropy:2013, astropy:2022}. The plots shown in this publication have been generated with the \textit{matplotlib} package \citep{Hunter:2007}. We furthermore made use of the Python packages \textit{numba}, \textit{numpy}, and \textit{pandas} \citep{2015_numba,harris2020array,McKinney_2011}.

\bibliographystyle{elsarticle-num}
\addcontentsline{toc}{part}{Bibliography}
\bibliography{adapt_kde}

\end{document}